\documentclass[final]{beatcs}
\usepackage{amsmath,amssymb,amsfonts,latexsym,stmaryrd}
\usepackage[PostScript=dvips]{diagrams}
\usepackage{euscript}
\usepackage{epsfig}
\usepackage{tikz,ifdraft}
\usetikzlibrary{arrows,positioning,matrix,backgrounds,calc,fit,decorations.pathreplacing} 

\usepackage{eurosym}

\newcommand{\beqa}{\begin{eqnarray*}}
\newcommand{\eeqa}{\end{eqnarray*}\par\noindent}

\newcommand{\vsa}{\vspace{.1in}}

\renewcommand{\emph}[1]{\textbf{#1}}

\newcommand{\lrarr}{\longrightarrow}
\newcommand{\rarr}{\rightarrow}

\newcommand{\ket}[1]{{|} #1\rangle}
\newcommand{\ie}{\textit{i.e.}~}

\newcommand{\IMP}{\; \rightarrow \;}

\newcommand{\da}{{\downarrow}}

\newcommand{\card}[1]{|#1|}

\newarrow{Eq}=====
\newarrow{mon}>--->
\newarrow{rel}--+->
\newarrow{inc}C--->
\newarrow{mto}|--->
\newarrow{LRTo}<--->
\newarrow{dash}....>

\diagramstyle[newarrowhead=vee,newarrowtail=vee]

\newcommand{\UU}{\EuScript{U}}

\newcommand{\Real}{\mathbb{R}}

\newcommand{\Bool}{\mathsf{B}}

\newcommand{\DB}{\mathcal{D}_{\Bool}}

\newcommand{\Up}{\mathsf{Up}}
\newcommand{\Down}{\mathsf{Down}}

\newcommand{\Complex}{\mathbb{C}}
\newcommand{\ua}{{\uparrow}}

\newcommand{\MM}{\mathcal{M}}

\newcommand{\prob}{\mathsf{Prob}}
\newcommand{\vphi}{\varphi}

\newcommand{\Att}{\mathcal{A}}

\renewcommand{\mid}{\, : \,}

\newcommand{\RS}{\mathcal{R}}
\newcommand{\nj}{\bowtie}
\newcommand{\SQT}{\!\sqrt{2}}
\newcommand{\ONE}{\mathsf{ONE}}

\title{Contextual Semantics:\\ From Quantum Mechanics to Logic, Databases, Constraints, and Complexity}

\author{Samson Abramsky\\
Department of Computer Science\\The University of Oxford\\\texttt{samson.abramsky@cs.ox.ac.uk}}

\date{}

\begin{document}

\beatcsColumn{logic}{}

\newpage
\maketitle

\begin{abstract}
We discuss quantum non-locality and contextuality, emphasising logical and structural aspects. We also show how the same mathematical structures arise in various areas of classical computation.
\end{abstract}

\section{Introduction}

In this paper we shall discuss some fundamental concepts in quantum mechanics: \emph{non-locality}, \emph{contextuality} and \emph{entanglement}. These concepts play a central r\^ole in the rapidly developing field of quantum information, in delineating how quantum resources can transcend the bounds of classical information processing. They also have profound consequences for our understanding of the very nature of physical reality.

Our aim is to present these ideas in a manner which should be accessible to any computer scientist, and which emphasises the logical and structural aspects. We shall also show how the same mathematical structures which arise in our analysis of these ideas  appear in a range of contexts in classical computation.

\section{Alice and Bob look at bits}

We consider the following scenario, depicted in Figure~1.
Alice and Bob are agents positioned at nodes of a network. Alice can access local bit registers $a_1$ and $a_2$, while Bob can access local bit registers $b_1$, $b_2$. Alice can load one of her bit registers into a processing unit, and test whether it is $0$ or $1$. Bob can perform the same operations with respect to his bit registers.
They send the outcomes of these operations to a common target, which keeps a record of the joint outcomes.

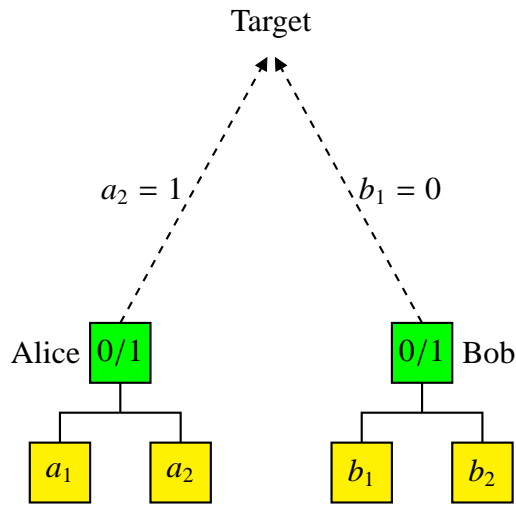
\begin{figure}
\label{bitreg}
\begin{center}
\begin{tikzpicture}[scale=2.54]
\ifx\dpiclw\undefined\newdimen\dpiclw\fi
\global\def\dpicdraw{\draw[line width=\dpiclw]}
\global\def\dpicstop{;}
\dpiclw=0.8bp
\dpicdraw[fill=green](0,-0.15625) rectangle (0.3125,0.15625)\dpicstop
\draw (0.15625,0) node{$0/1$};
\dpicdraw[fill=yellow](-0.3125,-0.78125) rectangle (0,-0.46875)\dpicstop
\draw (-0.15625,-0.625) node{$a_1$};
\dpicdraw[fill=yellow](0.3125,-0.78125) rectangle (0.625,-0.46875)\dpicstop
\draw (0.46875,-0.625) node{$a_2$};
\dpicdraw (-0.15625,-0.46875)
 --(-0.15625,-0.3125)
 --(0.46875,-0.3125)
 --(0.46875,-0.46875)\dpicstop
\dpicdraw (0.15625,-0.15625)
 --(0.15625,-0.3125)\dpicstop
\draw (0,0) node[left=-0.46875bp]{Alice};
\dpicdraw[fill=green](1.5625,-0.15625) rectangle (1.875,0.15625)\dpicstop
\draw (1.71875,0) node{$0/1$};
\dpicdraw[fill=yellow](1.25,-0.78125) rectangle (1.5625,-0.46875)\dpicstop
\draw (1.40625,-0.625) node{$b_1$};
\dpicdraw[fill=yellow](1.875,-0.78125) rectangle (2.1875,-0.46875)\dpicstop
\draw (2.03125,-0.625) node{$b_2$};
\dpicdraw (1.40625,-0.46875)
 --(1.40625,-0.3125)
 --(2.03125,-0.3125)
 --(2.03125,-0.46875)\dpicstop
\dpicdraw (1.71875,-0.15625)
 --(1.71875,-0.3125)\dpicstop
\draw (1.875,0) node[right=-0.46875bp]{Bob};
\draw (0.9375,1.59375) node[above=-0.46875bp]{Target};
\dpicdraw[dashed](0.15625,0.15625)
 --(0.885348,1.485604)\dpicstop
\filldraw[line width=0bp](0.912747,1.470576)
 --(0.915403,1.540403)
 --(0.857948,1.500631) --cycle
\dpicstop
\draw (0.534893,0.846625) node[left=-0.46875bp]{$a_2 = 1$};
\dpicdraw[dashed](1.71875,0.15625)
 --(0.989652,1.485604)\dpicstop
\filldraw[line width=0bp](1.017052,1.500631)
 --(0.959597,1.540403)
 --(0.962253,1.470576) --cycle
\dpicstop
\draw (1.340107,0.846625) node[right=-0.46875bp]{$b_1 = 0$};
\end{tikzpicture}
\end{center}
\caption{Alice and Bob look at bits}
\end{figure}

We now suppose that Alice and Bob perform repeated rounds of these operations. On different rounds, they may make different choices of which bit registers to access, and they may observe different outcomes for a given choice of register.
The target can compile statistics for this series of data, and infer probability distributions on the outcomes.
The probability table in Figure~2 records the outcome of such a process.

\begin{figure}
\label{probtab}
\begin{center}
\begin{tabular}{ll|ccccc}
A & B & $(0, 0)$ & $(1, 0)$ & $(0, 1)$ & $(1, 1)$  &  \\ \hline
$a_1$ & $b_1$ & $1/2$ & $0$ & $0$ & $1/2$ & \\
$a_1$ & $b_2$ & $3/8$ & $1/8$ & $1/8$ & $3/8$ & \\
$a_2$ & $b_1$ & $3/8$ & $1/8$ & $1/8$ & $3/8$ &  \\
$a_2$ & $b_2$ & $1/8$ & $3/8$ & $3/8$ & $1/8$ & 
\end{tabular}
\end{center}
\caption{The Bell table}
\end{figure}

Consider for example the cell at row 2, column 3 of the table. This corresponds to the following event:
\begin{itemize}
\item Alice loads register $a_1$ and observes the value $0$.
\item Bob loads register $b_2$ and observes the value $1$.
\end{itemize}
This event has the probability $1/8$, conditioned on Alice's choice of $a_1$ and Bob's choice of $b_2$.

Each row of the table specified a probability distribution on the possible joint outcomes, conditioned on the indicated choice of bit registers by Alice and Bob.

We can now ask:
\begin{center}
\fbox{How can such an observational scenario be realised?}
\end{center}

The obvious classical mechanism we can propose to explain these observations is depicted in Figure~3.

\begin{figure}
\label{sbitreg}
\begin{center}
\begin{tikzpicture}[scale=2.54]
\ifx\dpiclw\undefined\newdimen\dpiclw\fi
\global\def\dpicdraw{\draw[line width=\dpiclw]}
\global\def\dpicstop{;}
\dpiclw=0.8bp
\dpicdraw[fill=green](0,-0.094961) rectangle (0.227906,0.094961)\dpicstop
\draw (0.113953,0) node{$0/1$};
\dpicdraw[fill=yellow](-0.17093,-0.474804) rectangle (0.018992,-0.284883)\dpicstop
\draw (-0.075969,-0.379844) node{$a_1$};
\dpicdraw[fill=yellow](0.208914,-0.474804) rectangle (0.398836,-0.284883)\dpicstop
\draw (0.303875,-0.379844) node{$a_2$};
\dpicdraw (-0.075969,-0.284883)
 --(-0.075969,-0.189922)
 --(0.303875,-0.189922)
 --(0.303875,-0.284883)\dpicstop
\dpicdraw (0.113953,-0.094961)
 --(0.113953,-0.189922)\dpicstop
\draw (0,0) node[left=-0.284883bp]{Alice};
\dpicdraw[fill=green](0.987593,-0.094961) rectangle (1.215499,0.094961)\dpicstop
\draw (1.101546,0) node{$0/1$};
\dpicdraw[fill=yellow](0.816664,-0.474804) rectangle (1.006585,-0.284883)\dpicstop
\draw (0.911625,-0.379844) node{$b_1$};
\dpicdraw[fill=yellow](1.196507,-0.474804) rectangle (1.386429,-0.284883)\dpicstop
\draw (1.291468,-0.379844) node{$b_2$};
\dpicdraw (0.911625,-0.284883)
 --(0.911625,-0.189922)
 --(1.291468,-0.189922)
 --(1.291468,-0.284883)\dpicstop
\dpicdraw (1.101546,-0.094961)
 --(1.101546,-0.189922)\dpicstop
\draw (1.215499,0) node[right=-0.284883bp]{Bob};
\draw (0.60775,0.968601) node[above=-0.284883bp]{Target};
\dpicdraw[dashed](0.113953,0.094961)
 --(0.575484,0.903194)\dpicstop
\filldraw[line width=0bp](0.591977,0.893776)
 --(0.59432,0.936179)
 --(0.558992,0.912612) --cycle
\dpicstop
\draw (0.353552,0.514546) node[left=-0.284883bp]{$a_2 = 1$};
\dpicdraw[dashed](1.101546,0.094961)
 --(0.640015,0.903194)\dpicstop
\filldraw[line width=0bp](0.656508,0.912612)
 --(0.621179,0.936179)
 --(0.623522,0.893776) --cycle
\dpicstop
\draw (0.861948,0.514546) node[right=-0.284883bp]{$b_1 = 0$};
\dpicdraw[fill=lightgray](0.417828,-1.538366) rectangle (0.512789,-1.348445)\dpicstop
\draw (0.465308,-1.443405) node{$0$};
\dpicdraw[fill=lightgray](0.512789,-1.538366) rectangle (0.60775,-1.348445)\dpicstop
\draw (0.560269,-1.443405) node{$1$};
\dpicdraw[fill=lightgray](0.60775,-1.538366) rectangle (0.702711,-1.348445)\dpicstop
\draw (0.65523,-1.443405) node{$0$};
\dpicdraw[fill=lightgray](0.702711,-1.538366) rectangle (0.797671,-1.348445)\dpicstop
\draw (0.750191,-1.443405) node{$1$};
\dpicdraw[fill=lightgray](0.417828,-1.91821) rectangle (0.797671,-1.538366)\dpicstop
\draw (0.60775,-1.728288) node{$\vdots$};
\dpicdraw[dashed](0.465308,-1.348445)
 --(-0.055963,-0.507094)\dpicstop
\filldraw[line width=0bp](-0.039819,-0.497091)
 --(-0.075969,-0.474804)
 --(-0.072108,-0.517096) --cycle
\dpicstop
\dpicdraw[dashed](0.560269,-1.348445)
 --(0.314571,-0.511252)\dpicstop
\filldraw[line width=0bp](0.332795,-0.505903)
 --(0.303875,-0.474804)
 --(0.296348,-0.5166) --cycle
\dpicstop
\dpicdraw[dashed](0.65523,-1.348445)
 --(0.900928,-0.511252)\dpicstop
\filldraw[line width=0bp](0.919152,-0.5166)
 --(0.911625,-0.474804)
 --(0.882704,-0.505903) --cycle
\dpicstop
\dpicdraw[dashed](0.750191,-1.348445)
 --(1.271463,-0.507094)\dpicstop
\filldraw[line width=0bp](1.287607,-0.517096)
 --(1.291468,-0.474804)
 --(1.255318,-0.497091) --cycle
\dpicstop
\draw (0.60775,-1.91821) node[below=-0.284883bp]{Source};
\end{tikzpicture}
\end{center}
\caption{A Source}
\end{figure}
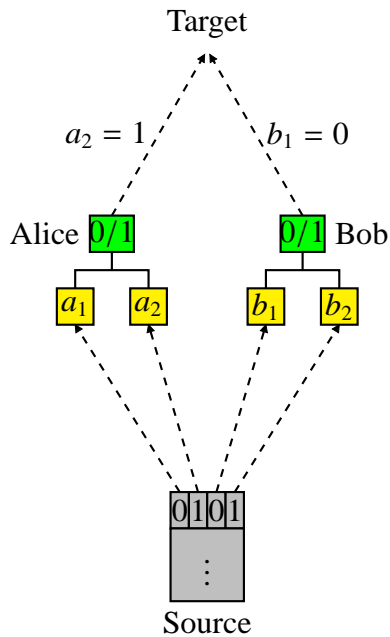
We postulate a \emph{source} which on each round chooses values for each of the registers $a_1$, $a_2$, $b_1$, $b_2$, and loads each register with the chosen value. Alice and Bob will then observe the values which have been loaded by the source.
We can suppose that this source is itself randomised, and chooses the values for the registers according to some probability distribution $P$ on the set of $2^{4}$ possible assignments.

We can now ask the question: is there any distribution $P$ which would give rise to the table specified in Figure~2?

\paragraph{Important Note}
A key observation is that, in order for this question to be non-trivial, we must assume that the choices of bit registers made by Alice and Bob are \emph{independent} of the source.\footnote{This translates formally into a conditional independence assumption, which we shall not spell out here; see e.g. \cite{Bellbeables,brandenburger2008classification}.} If the source  could determine which registers are to be loaded on each round, as well as their values, then it becomes a trivial matter to achieve \emph{any} given probability distribution on the joint outcomes.

Under this assumption of independence, it becomes natural to think of this scenario as a kind of \emph{correlation game}. The aim of the source is to achieve as high a degree of correlation between the outcomes of Alice and Bob as possible, whatever the choices made by Alice and Bob on each round.

\section{Logic rings a Bell}
We shall now make a very elementary and apparently innocuous deduction in elementary logic and probability theory, which could easily be carried out by students in the first few weeks of a Probability~$101$ course.

Suppose we have propositional formulas $\vphi_1, \ldots , \vphi_N$.
We suppose further that we can assign a probability $p_i$ to each $\vphi_i$.

In particular, we have in the mind the situation where the boolean variables appearing in $\vphi_i$ correspond to empirically testable quantities, such as the values of bit registers in our scenario; $\vphi_i$ then expresses a condition on the outcomes of an
experiment involving these quantities.
The probabilities $p_i$ are obtained from the statistics of these experiments.

Now suppose that these formulas are \emph{not simultaneously satisfiable}.  Then (e.g.)
\[ \bigwedge_{i=1}^{N-1} \phi_i \IMP \neg \phi_N ,  \quad \mbox{or equivalently} 
\quad \phi_N \IMP \bigvee_{i=1}^{N-1} \neg \phi_i . \]
Using elementary probability theory, we can calculate:
 \[ p_N  \; \leq \; \prob(\bigvee_{i=1}^{N-1} \neg \phi_i) \; \leq \; \sum_{i=1}^{N-1} \prob(\neg \phi_i) \; = \; \sum_{i=1}^{N-1} (1 - p_i) \; = \; (N - 1) - \sum_{i=1}^{N-1} p_i . \]
The first inequality is the monotonicity of probability, and the second is sub-additivity.

Hence we obtain the inequality
 \[ \sum_{i=1}^N p_i \; \leq \; N-1. \]
We shall refer to this as a \emph{logical Bell inequality}, for reasons to be discussed later. Note that it hinges on a purely logical consistency condition.

\subsection{Logical analysis of the Bell table}

We return to the probability table from Figure~2.

\begin{center}
\begin{tabular}{l|ccccc}
& $(0, 0)$ & $(1, 0)$ & $(0, 1)$ & $(1, 1)$  &  \\ \hline
$(a_1, b_1)$ & { \fcolorbox{gray}{yellow}{1/2}} & $0$ & $0$ & { \fcolorbox{gray}{yellow}{1/2}} & \\
$(a_1, b_2)$ &  { \fcolorbox{gray}{yellow}{3/8}} & $1/8$ & $1/8$ & { \fcolorbox{gray}{yellow}{3/8}} & \\
$(a_2, b_1)$ & { \fcolorbox{gray}{yellow}{3/8}} & $1/8$ & $1/8$ & { \fcolorbox{gray}{yellow}{3/8}} &  \\
$(a_2, b_2)$ & $1/8$ & { \fcolorbox{gray}{yellow}{3/8}} & { \fcolorbox{gray}{yellow}{3/8}} & $1/8$ & 
\end{tabular}
\end{center}

If we read $0$ as true and $1$ as false, the highlighted entries in each row of the table are represented by the following propositions:
\[ \begin{array}{rcccccccc}
\vphi_1 & = &  (a_{1} \wedge b_{1}) & \vee & (\neg a_{1} \wedge \neg b_{1}) & = & a_{1} & \leftrightarrow & b_{1} \\
\vphi_2 & = & (a_{1} \wedge b_{2}) & \vee & (\neg a_{1} \wedge \neg b_{2}) & = &  a_{1} & \leftrightarrow & b_{2} \\
\vphi_3 & = & (a_{2} \wedge b_{1}) & \vee & (\neg a_{2} \wedge \neg b_{1}) & = &  a_{2} & \leftrightarrow & b_{1} \\
\vphi_4 & = & (\neg a_{2} \wedge b_{2}) & \vee & (a_{2} \wedge \neg b_{2}) & = & a_{2} & \oplus & b_{2} .
\end{array}
\]
The events on first three rows are the correlated outcomes; the fourth is anticorrelated.
These propositions are easily seen to be jointly unsatisfiable. Indeed, starting with
$\vphi_4$, we can replace $a_{2}$ with $b_{1}$ using $\vphi_3$, $b_{1}$ with $a_{1}$ using $\vphi_1$, and $a_{1}$ with $b_{2}$ using $\vphi_2$, to obtain $b_{2} \oplus b_{2}$, which is obviously unsatisfiable.

It follows that our logical Bell inequality should apply, yielding the inequality
\[ \sum_{i=1}^4 p_i \; \leq \; 3. \]
However, we see from the table that $p_1 = 1$, $p_i = 6/8$ for $i = 2, 3, 4$.
Hence the table yields a violation of the Bell inequality by $1/4$.

This rules out the possibility of giving an explanation for the observational behaviour described by the table in terms of a classical source.
We might then conclude that such behaviour simply cannot be realised. However, as we shall now see, \emph{in the presence of quantum resources, this is no longer the case}.

\subsection{A crash course in qubits}

We shall now very briefly give enough information about some of the primitives of quantum computing to show how these can be used to realise behaviour such as that in the Bell table in Figure~2.
There are a number of excellent introductions to quantum computing aimed at or accessible to computer scientists (see e.g.~\cite{mermin2007quantum,yanofsky2008quantum,nielsen2000quantum}), and we refer the reader seeking more detailed information to these.

A classical bit register of the kind we began our discussion with can hold the values $0$ or $1$; we can say that it has two possible states.
The operations we can perform on such a register, or an array of such registers, include:
\begin{itemize}
\item Reading the value currently held in the register without changing the state of the register.
\item Using the values currently held in one or more such registers to compute a new value according to any boolean function.
\end{itemize}


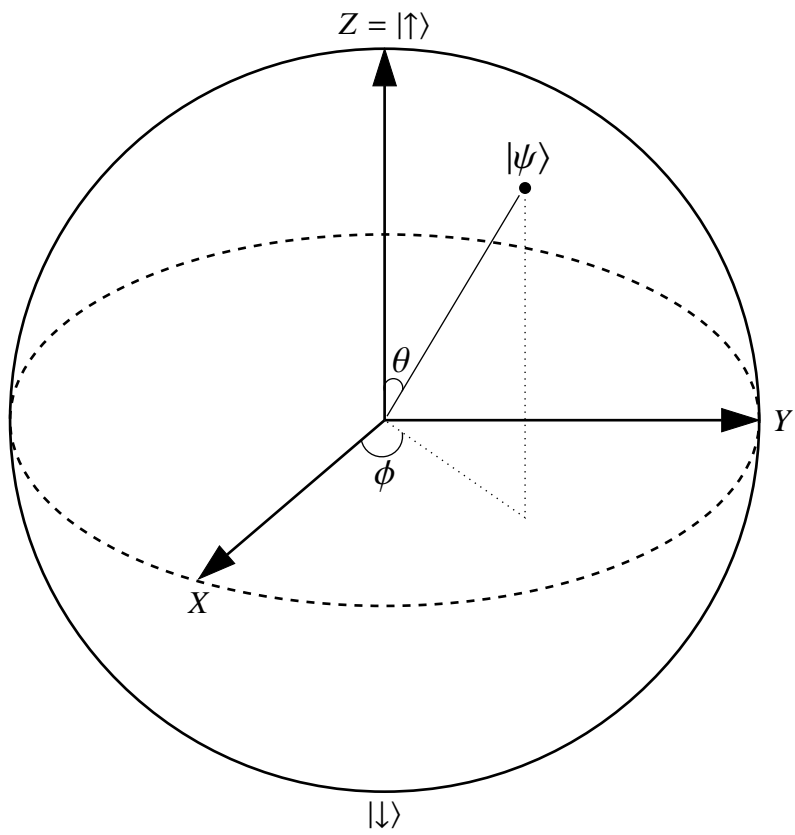
\begin{figure}
\label{Bloch}
\begin{center}
\begin{tikzpicture}[scale=2.54]
\ifx\dpiclw\undefined\newdimen\dpiclw\fi
\global\def\dpicdraw{\draw[line width=\dpiclw]}
\global\def\dpicstop{;}
\dpiclw=0.8bp
\dpiclw=1bp
\dpicdraw (1.939394,0) circle (0.763541in)\dpicstop
\dpicdraw[dashed](1.939394,0) ellipse (0.763541in and 0.38177in)\dpicstop
\dpicdraw (1.939394,0)
 --(1.939394,1.745455)\dpicstop
\filldraw (2,1.745455)
 --(1.939394,1.939394)
 --(1.878788,1.745455) --cycle
\dpicstop
\dpicdraw (1.939394,0)
 --(3.684848,0)\dpicstop
\filldraw (3.684848,-0.060606)
 --(3.878788,0)
 --(3.684848,0.060606) --cycle
\dpicstop
\dpicdraw (1.939394,0)
 --(1.123644,-0.702302)\dpicstop
\filldraw (1.084102,-0.656373)
 --(0.97667,-0.828837)
 --(1.163187,-0.748232) --cycle
\dpicstop
\dpicdraw[fill=black](2.666667,1.212121) circle (0.009544in)\dpicstop
\dpiclw=0.5bp
\dpicdraw (1.951779,0.02084)
 --(2.63714,1.174139)\dpicstop
\dpicdraw[dotted](2.666667,1.187879)
 --(2.666667,-0.509091)
 --(1.939394,0)\dpicstop
\dpicdraw (1.819053,-0.103605)
 ..controls (1.83009,-0.16194) and (1.886327,-0.200282)
 ..(1.944662,-0.189245)
 ..controls (2.002997,-0.178208) and (2.04134,-0.121971)
 ..(2.030303,-0.063636)\dpicstop
\dpicdraw (1.939394,0.151515)
 ..controls (1.927638,0.183187) and (1.951065,0.216872)
 ..(1.984848,0.216872)
 ..controls (2.018632,0.216872) and (2.042059,0.183187)
 ..(2.030303,0.151515)\dpicstop
\draw (2.683809,1.229263) node[above=-0.727273bp]{{\large $\ket{\psi}$}};
\draw (1.939394,-0.290909) node{{\large $\phi$}};
\draw (2.024242,0.315152) node{{\large $\theta$}};
\draw (1.939394,2.060606) node{$Z = \ket{{\uparrow}}$};
\draw (1.939394,-2.060606) node{$\ket{\da}$};
\draw (4,0) node{$Y$};
\draw (0.97667,-0.950049) node{$X$};
\end{tikzpicture}
\end{center}
\caption{The Bloch sphere representation of qubits}
\end{figure}

In quantum computing, we introduce a new object, the qubit, with very different properties. 
The key features of the qubit are best explained using the beautiful geometric representation  in terms of the ``Bloch sphere'' (the unit 2-sphere), as illustrated in Figure~4. 

Note the following key features:
\begin{itemize}
\item States of the qubit\footnote{More precisely, the pure states; mixed states are represented as points in the interior of the sphere.} are represented as points on the surface of the sphere. In Figure~4, a state $\ket{\psi}$ is depicted. Note that there are a continuum of possible states.
\item Each pair $(\Up, \Down)$ of antipodal points on the sphere define a possible measurement that we can perform on the qubit. Each such measurement has two possible outcomes, corresponding to $\Up$ and $\Down$ in the given direction. We can think of this physically e.g.~as measuring Spin Up or Spin Down in a given direction in space.
\item When we subject a qubit to a measurement $(\Up, \Down)$, the state of the qubit determines a probability distribution on the two possible outcomes. For a geometric view on this see Figure~5.
The probabilities are determined by the \emph{angles} between the qubit state $\ket{\psi}$ and the points $(\ket{{\Up}}, \ket{{\Down}})$ which specify the measurement. In algebraic terms, $\ket{\psi}$, $\ket{\Up}$ and $\ket{\Down}$ are unit vectors in the complex vector space $\Complex^2$, and the probability of observing $\Up$ when in state $\ket{\psi}$ is given by the square modulus\footnote{Recall that the square modulus of a complex number $z = a + ib$ is given by $| z |^2 = zz^* = (a + ib)(a - ib)$.} of the inner product:
\[ |\langle \psi | {\Up} \rangle |^2 . \]
This is known as the \emph{Born rule}. It gives the basic predictive content of quantum mechanics.
\item Note in addition that a measurement has an \emph{effect} on the state, which will no longer be the original state $\ket{\psi}$, but rather one of the states $\Up$ or $\Down$, in  accordance with the measured value.
\end{itemize}

\begin{figure}
\label{truthangle}
\begin{center}
\begin{tikzpicture}[scale=2.54]
\ifx\dpiclw\undefined\newdimen\dpiclw\fi
\global\def\dpicdraw{\draw[line width=\dpiclw]}
\global\def\dpicstop{;}
\dpiclw=0.8bp
\dpiclw=1bp
\dpicdraw[fill=yellow,draw=blue](1,0) circle (0.393701in)\dpicstop
\dpicdraw[fill=black](0.741181,0.965926) circle (0.009843in)\dpicstop
\dpicdraw[fill=black](1.258819,-0.965926) circle (0.009843in)\dpicstop
\dpicdraw[fill=black](1.707107,0.707107) circle (0.009843in)\dpicstop
\dpicdraw (0.741181,0.965926)
 --(1.258819,-0.965926)\dpicstop
\dpicdraw (1,0)
 --(1.707107,0.707107)\dpicstop
\dpiclw=2bp
\dpicdraw (0.967648,0.120741)
 ..controls (1.128635,0.163877) and (1.19334,-0.077604)
 ..(1.032352,-0.120741)\dpicstop
\dpicdraw (1.682959,0.713577)
 --(0.813626,0.946514)\dpicstop
\filldraw (0.820096,0.970663)
 --(0.765329,0.959455)
 --(0.807155,0.922366) --cycle
\dpicstop
\dpicdraw (1.700636,0.682959)
 --(1.27823,-0.893481)\dpicstop
\filldraw (1.254082,-0.887011)
 --(1.265289,-0.941778)
 --(1.302378,-0.899952) --cycle
\dpicstop
\draw (0.741181,0.990926) node[above=-0.75bp]{{$\ket{\Up}$}};
\draw (1.258819,-0.990926) node[below=-0.75bp]{{$\ket{\Down}$}};
\draw (1.724784,0.724784) node[right=-0.75bp]{{$\, {|} \psi \rangle$}};
\draw (1.075,0.3) node{$\theta^{\mathsf{U}}$};
\draw (1.2,-0.15) node{$\theta^{\mathsf{D}}$};
\end{tikzpicture}
\end{center}
\caption{Truth makes an angle with reality}
\end{figure}
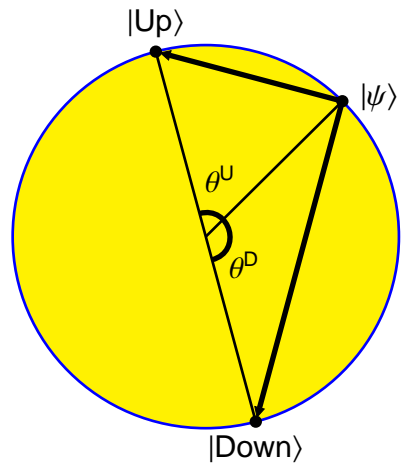

The sense in which the qubit generalises the classical bit is that, for each question we can ask --- \ie for each measurement --- there are just two possible answers. We can view the states of the qubit as superpositions of the classical states $0$ and $1$, so that we have a probability of getting each of the answers for any given state.

But in addition, we have the important feature that there are a continuum of possible questions we can ask.
However, note that on each run of the system, we can only ask \emph{one} of these questions. We cannot simultaneously observe $\Up$ or $\Down$ in two different directions.
Note that this corresponds to the feature of the scenario we discussed in Section~1, that Alice and Bob could only look at one their local registers on each round.

\subsection{Compound systems and entanglement}

The deeper features of quantum behaviour are revealed when we look at \emph{compound systems} of multiple qubits.\footnote{More generally, we can consider $d$-dimensional quantum systems for any positive integer $d$. A system of $n$ qubits has dimension $2^n$. Contextuality emerges already at dimension $d = 3$.}
It is here that we find the phenomena of quantum entanglement and non-locality.

\begin{figure}
\label{Bellstate}
\begin{center}
\begin{tikzpicture}[scale=2.54]
\ifx\dpiclw\undefined\newdimen\dpiclw\fi
\global\def\dpicdraw{\draw[line width=\dpiclw]}
\global\def\dpicstop{;}
\dpiclw=0.8bp
\dpicdraw[fill=white!10!black](0.1875,0) circle (0.073819in)\dpicstop
\dpicdraw[fill=white!10!black](2.8125,0) circle (0.073819in)\dpicstop
\dpicdraw (0.375,0)
 --(2.625,0)\dpicstop
\draw (1.5,0) node[above=-1.125bp]{$|\ua\ua \rangle + |\da\da\rangle$};
\end{tikzpicture}
\end{center}
\caption{The Bell state}
\end{figure}
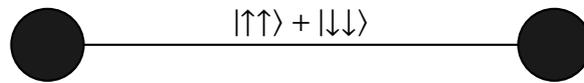

Consider for example the $2$-qubit system shown in Figure~6.
We can think of Alice holding one qubit, and Bob the other. The combined state of the system is described by the vector $\ket{\ua\ua} + \ket{\da\da}$.\footnote{We are ignoring normalisation constants.}
According to the standard postulates of quantum mechanics, when Alice measures her qubit, she may, with equal probability, get either answer (Spin Up or Down). If she gets the answer Spin Up, then the state of the entangled qubit becomes $\ket{\ua\ua}$, so that if Bob now measures his qubit, he can \emph{only} get the answer Spin Up; while if she gets the answer Spin Down, the state becomes $\ket{\da\da}$, and Bob can only get the answer Spin Down. This is regardless of the fact that Bob may be far away from Alice (spacelike separated). This is the phenomenon that Einstein famously referred to as ``spooky action at a distance'', and which Schr\"odinger named \emph{entanglement}.

How can the world be this way? This remains a challenge to our understanding of the nature of physical reality. Meanwhile, though, the field of quantum information seeks to understand how entanglement can be \emph{used} as a new kind of resource, opening up new possibilities which transcend those of the classical models of information and computation.

\subsection{From the Bell state to the Bell table}
We refer again to the table in Figure~2. This table can be physically realised, using the Bell state  $(\ket{\ua\ua} + \ket{\da\da})/\SQT$, with Alice and Bob performing 1-qubit local measurements corresponding to directions in the XY-plane of the Bloch sphere, at relative angle $\pi/3$.
Thus this behaviour is \emph{physically realisable} using quantum entanglement, although, as we have seen, it has no realisation by means of a classical source.

\subsubsection*{Computing the Bell table}

Some readers may find it helpful to see in detail how a table such as that in Figure~2 is computed. We shall now explain this.
Nothing following this subsection will depend on this material, so it can safely be skipped.

We shall consider spin measurements lying in the equatorial plane of the Bloch sphere, \ie the $XY$-plane as shown in Figure~4.
For such a measurement  at an angle $\phi$ to the $X$-axis, the Spin Up outcome is specified by the vector $({\ket{\ua} + e^{i\phi} \ket{\da}})/{\SQT}$, while the Spin Down outcome is specified by $({\ket{\ua} + e^{i(\phi + \pi)} \ket{\da}})/{\SQT}$. For the $X$ direction itself, we have $\phi = 0$, and these are the vectors $({\ket{\ua} + \ket{\da}})/{\SQT}$ and $({\ket{\ua} - \ket{\da}})/{\SQT}$ respectively.

We shall use the measurement in the $X$ direction for Alice's measurement $a_1$ and Bob's measurement $b_1$; while $a_2$ and $b_2$ will be interpreted by the measurements at angle $\phi = \pi/3$ to the $X$ axis. Note that Alice's measurements are applied to the first qubit of the Bell state, while Bob's measurements are applied to the second qubit.

For example, consider the following situation: Alice performs the measurement $a_1$ on the first qubit and observes the outcome $0$ (Spin Up), while Bob performs the measurement $b_2$ on the second qubit and observes outcome $1$ (Spin Down). This corresponds to the cell in row 2, column 3 of the table in Figure~2.
This event is represented by taking the tensor product of the vectors representing the outcomes for the local measurements by Alice and Bob on their qubits:
\[ \frac{\ket{\ua} + \ket{\da}}{\SQT} \otimes \frac{\ket{\ua} + e^{i4\pi/3} \ket{\da}}{\SQT} \;\; = \;\; \frac{\ket{\ua\ua} + e^{i4\pi/3} \ket{\ua\da} + \ket{\da\ua} + e^{i4\pi/3} \ket{\da\da}}{2} . \]
Call this vector $M$. The probability of observing this event when performing the joint measurement $(a_1, b_2)$ on the Bell state $B = (\ket{\ua\ua} + \ket{\da\da})/ \SQT$ is given, using the Born rule, by $| \langle B | M \rangle |^2$. Since the vectors $\ket{\ua\ua}$, $\ket{\ua\da}$, $\ket{\da\ua}$, $\ket{\da\da}$ are pairwise orthogonal, this simplifies to
\[ \left| \frac{1 + e^{i4\pi/3}}{2 \SQT} \right|^2 \;\; = \;\; \frac{| 1 + e^{i4\pi/3} |^2}{8} . \]
Using the Euler identity $e^{i\theta} = \cos \theta + i \sin \theta$, we have
\[ | 1 + e^{i\theta} |^2 \; = \; 2 + 2 \cos \theta . \]
Hence
\[ \frac{| 1 + e^{i4\pi/3} |^2}{8} \; = \; \frac{2 + 2 \cos (4\pi/3)}{8} \; = \; \frac{1}{8}, \]
the value given in the table in Figure~2. The other entries can be computed similarly.

\subsubsection*{Summary}
More broadly, we can say that this shows that quantum mechanics predicts correlations which exceed those which can be achieved by any classical mechanism. This is the content of \emph{Bell's theorem} \cite{bell1964einstein}, a famous result in the foundations of quantum mechanics, and in many ways the starting point for the whole field of quantum information.
Moreover, these predictions have been confirmed by many experiments which have been performed \cite{aspect1982experimental,aspect1999bell}.

\section{The ``Hardy Paradox''}

We shall now see how the same phenomena manifest themselves in a stronger form, which highlights a direct connection with logic.
Consider the table in Figure~7.

\begin{figure}
\label{Hardy}
\begin{center}
\begin{tabular}{c|cccc} 
 & $(0, 0)$ & $(0,1)$ & $(1,0)$ & $(1, 1)$ \\ \hline
$(a_1, b_1)$ &  $1$ &  &  &  \\
$(a_1, b_2)$ &   $0$ &  &  &  \\
$(a_2, b_1)$ &  $0$ &  &  & \\
$(a_2, b_2)$ &   &  &  & $0$ \\
\end{tabular}
\end{center}
\caption{The Hardy Paradox}
\end{figure}

This table depicts the same kind of scenario we considered previously. However, the entries are now either $0$ or $1$. The idea is that a $1$ entry represents a positive probability. Thus we are distinguishing only between \emph{possible} (positive probability) and \emph{impossible} (zero probability). In other words, the rows correspond to the \emph{supports} of some (otherwise unspecified) probability distributions.
Moreover, only four entries of the table are filled in. Our claim is that just from these four entries, referring only to the supports, we can deduce that there is no classical explanation for the behaviour recorded in the table. Moreover, this behaviour can again be realised in quantum mechanics, yielding a stronger form of Bell's theorem, due to Lucien Hardy \cite{hardy1993nonlocality}.\footnote{For a detailed discussion of  realisations of the Bell and Hardy models in quantum mechanics, see Section~7 of \cite{abramsky2013rhv}. Further details on the Hardy construction can be found in \cite{hardy1993nonlocality,mermin1994quantum}.}

\subsection{What Do ``Observables'' Observe?}

Classically, we would take the view that physical observables directly reflect properties of the physical system we are observing. These are objective properties of the system, which are independent of our choice of which measurements to perform ---
of our \emph{measurement context}.
More precisely, this would say that for each possible state of the system, there is a function $\lambda$ which for each measurement $m$ specifies an outcome $\lambda(m)$, \emph{independently of which other measurements may be performed}.
This point of view is called \emph{non-contextuality}, and may seem self-evident.
However, this view is \emph{impossible to sustain} in the light of our \emph{actual observations of (micro)-physical reality}.

Consider once again the Hardy table depicted in Figure~7. Suppose there is a function $\lambda$ which accounts for the possibility of Alice observing value $0$ for $a_1$ and Bob observing $0$ for $b_1$, as asserted by the entry in the top left position in the table. Then this function $\lambda$ must satisfy
 \[ \lambda : a_1 \mapsto 0, \quad b_1 \mapsto 0 . \]
Now consider the value of $\lambda$ at $b_2$. If $\lambda(b_2) = 0$, then this would imply that the event that $a_1$ has value $0$ and $b_2$ has value $0$ is possible. However, \emph{this is precluded} by the $0$ entry in the table for this event. The only other possibility is that $\lambda(b_2) = 1$. Reasoning similarly with respect to the joint values of $a_2$ and $b_2$, we conclude, using the bottom right entry in the table, that we must have $\lambda(a_2) = 0$. Thus the only possibility for $\lambda$ consistent with these entries is
\[ \lambda : a_1 \mapsto 0, \quad a_2 \mapsto 0, \quad b_1 \mapsto 0, \quad b_2 \mapsto 1 . \]
However, this would require the outcome $(0, 0)$ for measurements $(a_2,b_1)$ to be possible, and this is \emph{precluded} by the table.

We are thus forced to conclude that the Hardy models are contextual. Moreover, we can say that they are contextual in a logical sense, stronger than the probabilistic form we saw with the Bell tables, since we only needed information about possibilities to infer the contextuality of this behaviour.

\section{Mathematical Structure of Possibility Tables}

Consider again a table such as
\begin{center}
\begin{tabular}{l|cccc} 
 &  $(0, 0)$ & $(1, 0)$ & $(0, 1)$ & $(1, 1)$ \\ \hline
$(a_1, b_1)$ &  {{1}} &  $1$ &  $1$ &  $1$ \\
$(a_2, b_1)$ &   $0$ &  {{1}} &  {1} &  {1} \\
$(a_1, b_2)$ &  $0$ &  {1} &  {1} & {1} \\
$(a_2, b_2)$ &   {1} &  {1} &  {{1}} & $0$ \\
\end{tabular}
\end{center}

Let us anatomise the structure of this table.
There are  \emph{measurement contexts}  
\[ \{ a_1, b_1 \}, \quad \{ a_2, b_1 \}, \quad \{ a_1, b_2 \}, \quad \{ a_2, b_2 \} .\]
These are the possible combinations of measurements which can be made together, yielding the directly accessible empirical observations.\footnote{In quantum mechanics, these correspond to compatible families of observables.}
Each measurement has possible outcomes $0$ or $1$. More generally, we write $O$ for the set of possible outcomes.
Thus for example the matrix entry at row $(a_2, b_1)$ and column $(0,1)$ 
indicates the  \emph{event }
\[ \{ a_2 \mapsto 0,  \; b_1 \mapsto 1 \} . \]
The set of events relative to a context $C$ is the set of functions $O^C$.
Each row of the table specifies a \emph{Boolean distribution} on events $O^C$ for a given choice of measurement context $C$. Such a Boolean distribution is just a non-empty set of events.

Mathematically, this defines a \emph{presheaf}.
We have:

\begin{itemize}
\item  A set of measurements $X$ (the ``space''). In our example, $X = \{ a_1, a_2, b_1, b_2 \}$.
\item A family of subsets of $X$, the \emph{measurement contexts} (a ``cover''). In our example, these are
\[ \{ a_1, b_1 \}, \quad \{ a_2, b_1 \}, \quad \{ a_1, b_2 \}, \quad \{ a_2, b_2 \} \]
as already discussed.
\item To each such set $C$ a boolean distribution (finite non-empy subset) on \emph{local sections} $s : C \rarr O$, 
where $O$ is the set of \emph{outcomes}. Each row of the above table specifies such a distribution. Note that this notion of distribution generalises naturally to distributions valued in a \emph{commutative semiring}. We assume that the distributions have finite support, and are normalised (have total weight~$1$). In our case, we are using the idempotent semiring of the booleans.  We use the notation $\DB(X)$ for the set of boolean distributions on a set $X$.

Note that, if we use the semiring of non-negative reals instead, we obtain probability distributions with finite support.
\item A distribution on $C$ restricts to $C' \subseteq C$ by pointwise restriction of the local sections. More precisely, given such a distribution $d$ on $O^C$, we restrict it to $C'$ by defining, for $s \in O^{C'}$:
\[ d |_{C'} (s) \; = \; \sum_{s' \in O^C, s' |_{C'} = s} d(s') . \]
This definition makes sense for any semiring. In the boolean case, where addition is disjunction, it can be expressed equivalently as \emph{projection}, where we think of the distribution as a finite set:
\[ d |_{C'} \; = \; \{ s |_{C'} \; : \; s \in d \} . \]
In the probability case, it gives the usual notion of \emph{marginalisation}.
\end{itemize}

These local sections correspond to the directly observable \emph{joint outcomes} of \emph{compatible measurements}, which can actually be performed jointly on the system. 
The different sets of compatible measurements correspond to the different contexts of measurement and observation of the physical system.
The fact that the behaviour of these observable outcomes cannot be accounted for by some context-independent global description of reality corresponds to the geometric fact that these local sections cannot be glued together into a \emph{global section}.

For a picture of the familiar and simple situation of gluing functions together, consider the diagram in Figure~8.
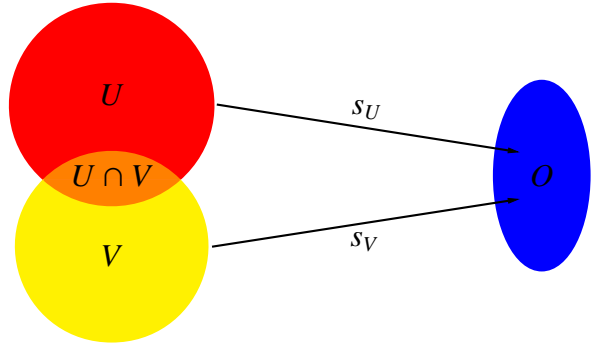
\begin{figure}
\label{gluefun}
\begin{center}
\begin{tikzpicture}[scale=2.54]
\ifx\dpiclw\undefined\newdimen\dpiclw\fi
\global\def\dpicdraw{\draw[line width=\dpiclw]}
\global\def\dpicstop{;}
\dpiclw=0.8bp
\dpicdraw[fill=red,line width=0.8bp,draw=red](0.144939,0.349915)
 ..controls (-0.215534,0.671387) and (0.011857,1.268011)
 ..(0.494854,1.268011)
 ..controls (0.977851,1.268011) and (1.205242,0.671387)
 ..(0.844769,0.349915)\dpicstop
\dpicdraw[fill=yellow,line width=0.8bp,draw=yellow](0.144939,0.349915)
 ..controls (-0.166801,0.038174) and (0.053986,-0.494854)
 ..(0.494854,-0.494854)
 ..controls (0.935722,-0.494854) and (1.156509,0.038174)
 ..(0.844769,0.349915)\dpicstop
\dpicdraw[fill=orange,line width=0.8bp,draw=orange](0.144939,0.349915)
 ..controls (0.338192,0.543167) and (0.651516,0.543167)
 ..(0.844769,0.349915)\dpicstop
\dpicdraw[fill=orange,line width=0.8bp,draw=orange](0.144939,0.349915)
 ..controls (0.34433,0.172097) and (0.645378,0.172097)
 ..(0.844769,0.349915)\dpicstop
\dpicdraw[fill=blue,draw=blue](2.721697,0.371141) ellipse (0.097412in and 0.194824in)\dpicstop
\dpicdraw (1.039194,0.742281)
 --(2.54911,0.502612)\dpicstop
\filldraw[line width=0bp](2.547171,0.490393)
 --(2.597984,0.494854)
 --(2.55105,0.51483) --cycle
\dpicstop
\draw (1.812991,0.619456) node[above=-0.742281bp]{$s_U$};
\dpicdraw (1.014451,0)
 --(2.549092,0.239788)\dpicstop
\filldraw[line width=0bp](2.551002,0.227565)
 --(2.597984,0.247427)
 --(2.547182,0.252011) --cycle
\dpicstop
\draw (1.800618,0.122839) node[below=-0.742281bp]{$s_V$};
\draw (0.494854,0.791767) node{$U$};
\draw (0.494854,-0.049485) node{$V$};
\draw (0.494854,0.371141) node{$U \cap V$};
\draw (2.721697,0.371141) node{$O$};
\end{tikzpicture}
\end{center}
\caption{Gluing functions}
\end{figure}
If $s_U |_{U \cap V} = s_V |_{U \cap V}$, they can be glued to form
\[ s : U \cup V \lrarr O \]
such that $s |_U = s_U$ and $s |_V = s_V$.

In geometric language, the Hardy paradox corresponds to the fact that there is a \emph{local section} which cannot be extended to a \emph{global section} which is compatible with the family of boolean distributions.
In other words, the space of \emph{local possibilities} is sufficiently logically `twisted' to \emph{obstruct} such an extension.

The quantum phenomena of \emph{non-locality} and \emph{contextuality} correspond exactly to the existence of obstructions to global sections in this sense.
This geometric language is substantiated by the results in \cite{abramsky2011cohomology}, which show that \emph{sheaf cohomology} can be used to characterise these obstructions, and to witness contextuality in a wide range of cases.

This geometric picture and the associated methods can be applied to a wide range of situations in classical computer science, which do not seem to have anything in common with the quantum realm.
In particular, as we shall now see, there is an isomorphism between the formal description we have given for the quantum notions of non-locality and contextuality, and basic definitions and concepts in relational database theory.

\section{Relational Databases and Bell's Theorem}

Consider an example of a table in a relational database, as in Figure~9.
\begin{figure}
\label{reltable}
\begin{center}
\begin{tabular}{|l|l|l|l|} \hline
\textbf{branch-name} & \textbf{account-no} & \textbf{customer-name} & \textbf{balance} \\ \hline \hline
Cambridge & $10991$-$06284$ & Newton & \pounds 2,567.53 \\ \hline
Hanover & $10992$-$35671$ & Leibniz & \euro 11,245.75 \\ \hline
\ldots & \ldots & \ldots & \ldots \\ \hline
\end{tabular}
\end{center}
\caption{A relation table}
\end{figure}

Let us anatomise such tables:
\begin{itemize}
\item The columns are determined by a set $A$ of \emph{attributes}. Assume $A \subset \Att$ for some global set $\Att$ specified by the database schema.
\item For each attribute $a$, there is a possible set of \emph{data values} $D_a$. For simplicity, we collect these  into a global set $D = \bigsqcup_{a \in \Att} D_a$.
\item An   \emph{A-tuple} is specified by a function $t : A \rarr D$. 
\item A \emph{relation instance} or \emph{table} of schema $A$ is a set of $A$-tuples.
\item A \emph{database schema} is given by a family  $\Sigma = \{A_1, \ldots , A_k\}$ of finite subsets of $\Att$.
\item A database \emph{instance} of schema $\Sigma$ is given by a family of relation instances $\{ R_i \}$ where $R_i$ is of schema $A_i$.
\end{itemize}

Does this look familiar? In fact, it is straightforward to express this structure in the language of presheaves:
\begin{itemize}
\item An $A$-tuple $t$ is just a local section over $A$: $t \in D^A$.
\item A relation table $R$ of  schema $A$ is a boolean distribution on $A$-tuples:
\[ R \in \DB (D^A) . \]
\item Note that if $A \subseteq B$, then restriction is just \emph{projection}. For $R \in \DB(D^B)$
\[ R |_A := \{ t |_A \mid t \in R \} . \]
\item We can regard a schema $\Sigma$ as a cover of $\Att$.
\item A database instance of schema $\Sigma$ is a family of elements $\{ R_A \}_{A \in \Sigma}$.
\item The compatibility condition for an instance is \emph{projection consistency}:
\[ R_A |_{A \cap B} = R_B |_{A \cap B} \]
means that the two relations have the same projections onto their common set of attributes.
\end{itemize}

\subsection{Universal Relations}

A \emph{universal relation} for an instance $\{ R_A \mid A \in \Sigma \}$ of a schema $\Sigma$ is a relation $R \in \DB(D^{\Att})$ such that, for all $A \in \Sigma$:
\[ R |_A = R_A . \]
Thus it is a relation defined on the whole set of attributes $\Att$ from which each of the relations in the instance can be recovered by projection.

This notion, and various related ideas,  played an important r\^ole in early developments in relational database theory; see e.g. \cite{maier1984foundations,fagin1982simplied,korth1984system,maier1983maximal,ullman1983principles}.
Note that a universal relation instance corresponds exactly to the notion of \emph{global section} for the database instance viewed as a compatible  family. (Compatibility is obviously a necessary condition for such an instance to exist).

It is also standard that a universal relation need not exist in general, and even if it exists, it need not be unique.
There is a substantial literature devoted to the issue of finding conditions under which these properties do hold.

There is a simple connection between universal relations and lossless joins.

\vsa
\textbf{Proposition}.
Let $(R_1, \ldots , R_k)$ be an instance for the schema $\Sigma = \{ A_1, \ldots , A_k \}$. Define $R := \; \nj_{i=1}^k R_i$.
Then a universal relation for the instance exists if and only if $R |_{A_i} = R_i$, $i = 1, \ldots , k$, and in this case $R$ is the largest relation in $\RS(\bigcup_i A_i)$ satisfying the condition for a global section. $\Box$

\vsa
We can summarise the striking correspondence we have found between the realms of quantum contextuality and database theory in the following dictionary:
\begin{center}
\begin{tabular}{l|l} 
Relational databases & measurement scenarios \\ \hline
attribute & measurement \\
set of attributes defining a relation table & compatible set of measurements \\
database schema & measurement cover \\
tuple & local section (joint outcome) \\
relation/set of tuples & boolean distribution on joint outcomes \\
universal relation instance & global section/hidden variable model \\
acyclicity & Vorob'ev condition \cite{vorob1962consistent}
\end{tabular}
\end{center}

This dictionary goes beyond what we have discussed so far. The last entry concerns Vorob'ev's Theorem \cite{vorob1962consistent}, a remarkable result motivated by game theory which provides a necessary and sufficient combinatorial condition on a set cover or hypergraph (formulated equivalently in terms of abstract simplicial complexes) such that any compatible family of probability distributions over this cover can be glued together into a global section --- a joint distribution on the whole set of vertices which marginalises to yield the given distribution over each simplex. This condition is equivalent to the well-studied notion of \emph{acyclicity} of database schemes \cite{beeri1983desirability,maier1983theory}.

It seems that there is considerable scope for taking these connections and common structures further. For example, we can consider probabilistic databases, and more generally distributions valued in semirings. See \cite{abramsky2013relational} for a more detailed discussion.

\subsection{Hidden variables and all that}
We mentioned hidden variable models in the above table, but have not otherwise done so in this article.
Traditionally, such models have played a leading r\^ole in discussions of quantum non-locality and contextuality.
Essentially, a local hidden-variable model is what we called a ``classical source'' in Section~3. Indeed, a standard way of picturing such a model, due to David Mermin \cite{mermin1990quantum}, is shown in Figure~10.
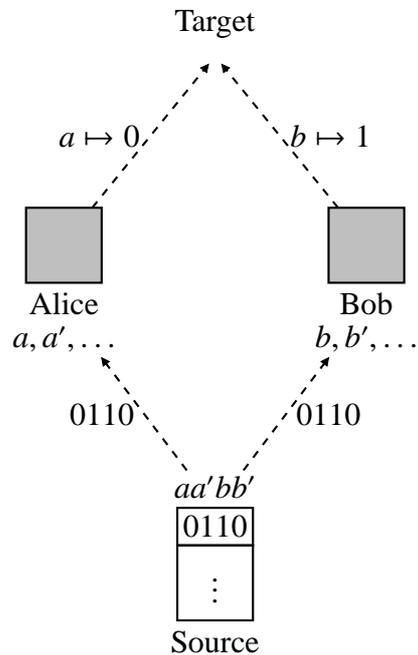
\begin{figure}
\label{sbitreg}
\begin{center}
\begin{tikzpicture}[scale=2.54]
\ifx\dpiclw\undefined\newdimen\dpiclw\fi
\global\def\dpicdraw{\draw[line width=\dpiclw]}
\global\def\dpicstop{;}
\dpiclw=0.8bp
\dpicdraw[fill=lightgray](0,-0.19583) rectangle (0.391659,0.19583)\dpicstop
\dpicdraw[fill=lightgray](1.566636,-0.19583) rectangle (1.958296,0.19583)\dpicstop
\draw (0.19583,-0.19583) node[below=-0.293744bp]{Alice};
\draw (1.762466,-0.19583) node[below=-0.293744bp]{Bob};
\draw (0.19583,-0.489574) node{$a, a', \ldots$};
\draw (1.762466,-0.489574) node{$b, b', \ldots$};
\dpicdraw (0.783318,-1.566636) rectangle (1.174977,-1.370807)\dpicstop
\draw (0.979148,-1.468722) node{$0110$};
\dpicdraw (0.783318,-1.958296) rectangle (1.174977,-1.566636)\dpicstop
\draw (0.979148,-1.762466) node{$\vdots$};
\draw (0.979148,-1.370807) node[above=-0.293744bp]{$aa'bb'$};
\draw (0.979148,-1.958296) node[below=-0.293744bp]{Source};
\dpicdraw[dashed](0.832276,-1.174977)
 --(0.415159,-0.618821)\dpicstop
\filldraw[line width=0bp](0.430825,-0.607072)
 --(0.391659,-0.587489)
 --(0.399492,-0.630571) --cycle
\dpicstop
\draw (0.612697,-0.882206) node[left=-0.293744bp]{$0110$};
\dpicdraw[dashed](1.12602,-1.174977)
 --(1.543137,-0.618821)\dpicstop
\filldraw[line width=0bp](1.558803,-0.630571)
 --(1.566636,-0.587489)
 --(1.527471,-0.607072) --cycle
\dpicstop
\draw (1.345598,-0.882206) node[right=-0.293744bp]{$0110$};
\draw (0.979148,1.028105) node[above=-0.293744bp]{Target};
\dpicdraw[dashed](0.342702,0.19583)
 --(0.919992,0.914003)\dpicstop
\filldraw[line width=0bp](0.935255,0.901734)
 --(0.94453,0.94453)
 --(0.904729,0.926272) --cycle
\dpicstop
\draw (0.642854,0.569232) node[left=-0.293744bp]{$a \mapsto 0$};
\dpicdraw[dashed](1.615594,0.19583)
 --(1.038304,0.914003)\dpicstop
\filldraw[line width=0bp](1.053567,0.926272)
 --(1.013766,0.94453)
 --(1.023041,0.901734) --cycle
\dpicstop
\draw (1.315442,0.569232) node[right=-0.293744bp]{$b \mapsto 1$};
\end{tikzpicture}
\end{center}
\caption{The Mermin instruction set picture}
\end{figure}
This is essentially the same picture as Figure~3. 
Mermin calls the hidden variables ``instruction sets''; these correspond exactly to the global assignments we have been discussing, which can be considered as canonical forms of hidden variables. It is shown in \cite[Theorem 8.1]{abramsky2011unified} that these are equivalent to the more general forms of hidden variable models which have been considered in the literature.

\subsection{Contextual semantics}
Why do such similar structures arise in such apparently different settings?
The phenomenon of contextuality is pervasive. Once we start looking for it, we can find it everywhere!
Examples already considered include: physics \cite{abramsky2011unified}, computation \cite{abramsky2013robust},  and natural language \cite{abramsky2014semantic}.

This leads to what we may call the  \emph{Contextual semantics hypothesis}: we can find common mathematical structure in all these diverse manifestations, and develop a widely applicable theory.

\section{Kochen-Specker Models}

We now return to quantum mechanics, and discuss another fundamental result, the Kochen-Specker theorem \cite{kochen1975problem}.\footnote{Since Bell independently proved a version of this result \cite{bell1966problem}, it is often called the Bell-Kochen-Specker theorem.}
This result shows the contextuality of quantum mechanics in an even stronger form than Bell's theorem, in the sense that the argument is independent of any particular quantum state. Whereas our arguments for the Bell and Hardy theorems hinged on realising contextual behaviours using certain entangled quantum states, the Kochen-Specker argument rests on properties of certain families of measurements which hold for \emph{any} quantum state.

There is, however, a trade-off. Whereas the conclusion of the Kochen-Specker theorem is stronger than that of Bell's theorem, its assumptions are also stronger, in that it assumes (for a contradiction) non-contextuality for measurements \emph{in general}. By contrast, Bell's theorem applies to a particular class of measurement scenarios where Alice and Bob are spacelike separated; in these situations, the assumption of non-contextuality is supported by relativistic considerations, which imply that there can be no direct causal influence by the measurements on each other.

The stronger form of state-independent contextuality given by the Kochen-Specker theorem is nevertheless of great interest, and has been the subject of a number of recent experimental verifications \cite{bartosik2009experimental,kirchmair2009state}. It is also a topic of current interest to develop methods for exploiting contextuality as a resource in quantum information, extending what has been done for non-locality.
A feature of our sheaf-theoretic framework, as described in Section~5, is that it provides a unified setting for Bell's theorem, the Kochen-Specker theorem, and other results relating to non-locality and contextuality.

We recall the general setting discussed in Section~5. We have a set $X$ of measurement labels, and a family $\UU$ of subsets of $X$ --- a ``measurement cover''. The sets $C \in \UU$ are the \emph{measurement contexts}; those combinations of measurements which can be performed together. Formally speaking, $(X, \UU)$ is just a hypergraph.

For convenience we fix our set of outcomes as $O = \{ 0, 1 \}$. Given $C \in \UU$, we say that $s \in O^C$ satisfies the \emph{KS property} if $s(x) = 1$ for exactly one $x \in C$. The \emph{Kochen-Specker model} over $(X, \UU)$ is defined by setting $d_C$, for each $C \in \UU$, to be the set of all $s \in O^C$ which satisfy the KS property. Note that the model is uniquely determined once we have given $(X, \UU)$.

Note that, if we regard the elements of $X$ as propositional variables, we can think of $s \in O^C$ as a truth-value assignment.\footnote{Interpreting $1$ as true and $0$ as false.}
Then the KS property for an assignment $s$ is equivalent to $s$ satisfying the following formula:
\[ \ONE(C) \; := \; \bigvee_{x \in C} (x \; \wedge \; \bigwedge_{x' \in C \setminus \{ x \}} \neg x') \]
We say that the Kochen-Specker model over $(X, \UU)$ is contextual if there is no global assignment $s : X \rarr O$ on the whole set of variables $X$ such that $s |_C \in d_C$ for all $C \in \UU$. Equivalently, we can say that the model is contextual if the formula
\[ \bigwedge_{C \in \UU} \ONE(C) \]
is unsatisfiable.\footnote{Note that in the general case where $O$ is some finite set, this becomes a constraint satisfaction problem. Contextuality means that the problem has no solution.}

It is interesting to compare this with the property of the Hardy models discussed in Section~5. As we saw there, the contextuality property exhibited by these models was that there was a local section in the support at some $C \in \UU$ which was not extendable to a global assignment on $X$ which was compatible with the support. By contrast, the form of contextuality we are considering here is much stronger; that \emph{there is no global assignment at all} which is consistent with the support. In fact, the Hardy models do not satisfy this stronger property. 

The simplest example of a contextual Kochen-Specker model is the triangle, \ie the cover
\[  \{a, b\}, \{ b, c \}, \{ a, c \}   \]
on $X = \{ a, b, c \}$.
For a more elaborate example, consider the set $X = \{ m_1, \ldots , m_{18} \}$, and the measurement  cover $\MM$ whose elements are the columns of the following table:
\begin{center}
\begin{tabular}{|c|c|c|c|c|c|c|c|c|} \hline
$m_1$ & $m_1$ & $m_8$ & $m_8$ & $m_2$ & $m_9$ & $m_{16}$ & $m_{16}$  & $m_{17}$ \\ \hline
$m_2$ & $m_5$ & $m_9$ & $m_{11}$ & $m_5$ & $m_{11}$ & $m_{17}$ & $m_{18}$ & $m_{18}$  \\ \hline
$m_3$ & $m_6$ & $m_3$ & $m_7$ & $m_{13}$ &  $m_{14}$ & $m_4$ & $m_6$ & $m_{13}$  \\ \hline
$m_4$ & $m_7$ & $m_{10}$ & $m_{12}$ & $m_{14}$  & $m_{15}$ & $m_{10}$ & $m_{12}$ & $m_{15}$  \\ \hline
\end{tabular}
\end{center}

How we do we show that a model such as this is contextual?
We shall give a combinatorial criterion on $(X, \UU)$ which can be used for most of the examples which have appeared in the literature.

For each $x \in X$, we define
\[ \UU(x) := \{ C \in \UU \mid x \in C \} . \]

\textbf{Proposition} \cite[Proposition 7.1]{abramsky2011unified}.
If the Kochen-Specker model on $(X, \UU)$ is non-contextual, then every common divisor of $\{ \card{\UU(x)} \mid x \in X \}$ must divide $\card{\UU}$. $\Box$

\vsa
Applying this to the above example, we note that the cover $\MM$ has 9 elements, while each element of $X$ appears in two members of $\MM$. Thus the Kochen-Specker model on $(X, \MM)$ is contextual.

\subsection*{Quantum representations}
What do these combinatorial questions have to do with quantum mechanics? A contextual Kochen-Specker model $(X, \UU)$ gives rise to a quantum mechanical witness of contextuality whenever we can label $X$ with unit vectors in $\Real^n$, for some fixed $n$, such that $\UU$ consists exactly of those subsets $C$ of $X$ which form orthonormal bases of $\Real^n$. The point of our example $\MM$ above is that it is possible to label the 18 elements of $X$ with vectors in $\Real^4$ such that the four-element subsets in $\MM$ are orthogonal \cite{cabello1996bell}. This yields one of the most economical known quantum witnesses for contextuality.\footnote{By contrast, the triangle does \emph{not} yield a quantum witness, since orthogonality is a pairwise notion; if all the pairs are orthogonal, the whole set must be also.}

To connect this directly to quantum measurement, note that such a family of vectors can be used to define corresponding measurements, such that the measurements corresponding to orthogonal sets are compatible, and moreover for any quantum state $\ket{\psi}$, the support of the distribution on outcomes induced by performing this joint measurement on $\ket{\psi}$ will satisfy the KS property. Thus contextuality of the model yields a state-independent witness of quantum contextuality. For a detailed discussion of this point, see Section~9.2 of \cite{abramsky2011unified}.

The smallest dimension for which contextuality witnesses appear in this form is $n=3$. Currently, the smallest known Kochen-Specker model providing a contextuality witness in dimension 3 has 31 vectors \cite{peres1995quantum}. Computational methods have established a lower bound of 18 \cite{arends2011searching}.

\section{Discussion and Further Reading}

One aim of this paper has been to present some central concepts of quantum information and foundations in a form which will be accessible to computer scientists, in particular those with an interest in logical and structural methods. At the same time, we have also aimed to provide an introduction to recent research by the author and a number of colleagues, which aims to use tools which have been developed within computer science logic and semantics to study these quantum notions. This ``high-level'' approach has led to a number of developments, both within quantum information, and in identifying the same formal structures in a number of classical computational situations; we have seen an example of this in the case of relational database theory.

We shall conclude by discussing some references where the interested reader can find further information, and see these ideas developed in greater depth.\footnote{The papers by the author which are referenced can be found at \texttt{arXiv.org}.}

\subsection{The sheaf-theoretic approach}

As discussed briefly in Section~5, our analysis of non-locality and contextuality uses the mathematical framework of sheaves and presheaves.
The issue of finding ``local realistic'' explanations of correlated behaviour is interpreted geometrically in terms of finding global sections in the sense of sheaf theory. These ideas, and many basic results, are developed in the paper \cite{abramsky2011unified} with Adam Brandenburger which laid the basis for this approach.

This leads to a number of developments in quantum information and foundations:
\begin{itemize}
\item The sheaf-theoretic language allows a unified treatment of non-locality and contextuality, in which results such as Bell's theorem \cite{bell1964einstein} and the Kochen-Specker theorem \cite{kochen1975problem} fit as instances of more general results concerning obstructions to global sections. In recent work \cite{mansfield2014extendability}, it has been shown how this framework can be used to \emph{transform} contextuality scenarios into non-locality scenarios.

\item A hierarchy of degrees of non-locality or contextuality is identified in \cite{abramsky2011unified}. This explains and generalises the notion of ``inequality-free'' or ``probability-free'' non-locality proofs, and makes a strong connection to logic, as developed in \cite{abramsky2013rhv}.
This hierarchy is lifted to a novel classification of multipartite entangled states, leading to some striking new results concerning multipartite entanglement, which is currently poorly understood. These results will appear in forthcoming joint publications with Carmen Constantin and Shenggang Ying.

\item The obstructions to global sections witnessing contextuality are characterised in terms of sheaf cohomology in \cite{abramsky2011cohomology} with Shane Mansfield and Rui Barbosa, and a range of examples are treated in this fashion.

\item A striking connection between no-signalling models and global sections with signed measures (``negative probabilities'') is established in \cite{abramsky2011unified}. An operational interpretation of such negative probabilities, involving a signed version of the strong law of large numbers, is developed in \cite{abramsky2014operational}.
\end{itemize}

\subsection{Logical Bell inequalities}

The discussion in Section~3 is based on \cite{abramsky2012logical}.
Bell inequalities are a central technique in quantum information. In \cite{abramsky2012logical} with Lucien Hardy, a general notion of ``logical Bell inequality'', based on purely logical consistency conditions, is introduced, and it is shown that every Bell inequality (\ie every inequality satisfied by the ``local polytope'') is equivalent to a logical Bell inequality. The notion is developed at the level of generality of  \cite{abramsky2011unified}, and hence applies to arbitrary contextuality scenarios, including multipartite Bell scenarios and Kochen-Specker configurations.

\subsection{Contextual semantics in classical computation}

We discussed the isomorphism between the basic concepts of quantum contextuality and those of relational database theory in Section~6.
A number of other connections have been studied:
\begin{itemize}
\item In \cite{abramsky2013rhv} connections between non-locality and logic are emphasised. A number of natural complexity and decidability questions are raised in relation to non-locality.
\item Our discussion of the Hardy paradox in Section~5 showed that the key issue was that a local section (assignment of values) could not be extended to a global one consistently with some constraints (the ``support table''). This directly motivated some joint work with Georg Gottlob and Phokion Kolaitis \cite{abramsky2013robust}, in which we studied a refined version of \emph{constraint satisfaction}, dubbed ``robust constraint satisfaction'', in which one asks if a partial assignment of a given length can always be extended to a solution. The tractability boundary for this problem is delineated in \cite{abramsky2013robust}, and this is used to settle one of the complexity questions posed in \cite{abramsky2013rhv}.
\item Application of the contextual semantics framework to natural language semantics is initiated in \cite{abramsky2014semantic} with Mehrnoosh Sadrzadeh. In this paper, a basic part of the Discourse Representation Structure framework \cite{kamp1993discourse} is formulated as a presheaf, and the gluing of local sections into global ones is used to represent the resolution of anaphoric references.
\end{itemize}

Further connections and applications of contextual semantics are currently being studied, and it seems likely that more will be forthcoming.

\bibliographystyle{plain} 

\bibliography{bdbib}
\end{document}